# Single-crystal structural phase diagram of stoichiometric bilayer nickelate $La_3Ni_2O_7$ under hydrostatic pressure

Misaki Sasaki[1,†], Zhehong Liu[2,†], Takeshi Hara[3,†], Shunsuke Kitou[4,*], Markus Kriener[2], Haruto Yoshimochi[2], Shion Yamada[4], Chieko Terakura[2], Naohisa Hirao[5], Hirokazu Kadobayashi[5], Yusuke Wakabayashi[3], Yoshinori Tokura[1,2,6], Yasujiro Taguchi[2], Takahisa Arima[2,4], Yukako Fujishiro[1,2,7,8,*]

[1]*Department of Applied Physics, The University of Tokyo, Tokyo 113-8654 Japan*

[2]*RIKEN Center for Emergent Matter Science (CEMS), Wako 351-0198, Japan*

[3]*Department of Physics, Graduate School of Science, Tohoku University, Sendai 980-8578, Japan*

[4]*Department of Advanced Materials Science, The University of Tokyo, Kashiwa 277-8561, Japan*

[5]*Japan Synchrotron Radiation Research Institute (JASRI), SPring-8, Hyogo 679-5198, Japan*

[6]*Tokyo College, The University of Tokyo, Tokyo 113-8654 Japan*

[7]*Institute of Engineering Innovation, The University of Tokyo, Tokyo 113-8656, Japan*

[8]*RIKEN Pioneering Research Institute (PRI), Saitama 351-0198, Japan*

*Corresponding authors: kitou@edu.k.u-tokyo.ac.jp; fujishiro@ap.t.u-tokyo.ac.jp

[†]These authors contributed equally

## Abstract

The bilayer nickelate $La_3Ni_2O_7$ has attracted intense interest following the discovery of high-temperature superconductivity under pressure, representing the first nickelate superconductor realized in bulk form [1]. However, the crystal structure of the superconducting phase remains under active discussion [2–5], complicating efforts to establish its microscopic origin. Here we resolve these structural controversies by establishing a definitive pressure–temperature phase diagram, including the superconducting region of stoichiometric $La_3Ni_2O_7$ single crystals under hydrostatic conditions using helium as the pressure-transmitting medium. At ambient pressure, $La_3Ni_2O_7$ adopts a polar orthorhombic *Am*2*m* structure characterized by charge order between inequivalent Ni sites and $NiO_6$ octahedral tilting. Upon compression, the system undergoes a direct transition from the charge-ordered *Am*2*m* phase to the tetragonal *I4/mmm* phase near 10 GPa, coinciding with the onset of bulk superconductivity. These results establish the intrinsic structural evolution of $La_3Ni_2O_7$ and provide a structural framework for microscopic theories of nickelate superconductivity.

## Introduction

The realization of high-temperature superconductivity in nickelates [1–21] has opened a new frontier for exploring unconventional superconductivity in correlated transition-metal oxides [22–23]. Following the discovery of superconductivity in infinite-layer nickelate thin films [6–10], the observation of a superconducting transition temperature, $T_C$, approaching 80 K in the bilayer Ruddlesden–Popper nickelate $La_3Ni_2O_7$ under high pressure [1] has stimulated intense experimental and theoretical activities [2–5,11–16]. This interest has been further amplified by recent reports of enhanced bulk superconductivity through chemical-pressure engineering [17–19], as well as by studies of related nickelates such as trilayer $La_4Ni_3O_{10}$ [20] and hybrid $La_5Ni_3O_{11}$ [21].

In contrast to the predominantly single-band electronic structure of the cuprates [24], $La_3Ni_2O_7$, with a formal Ni valence of 2.5+ ($3d^{7.5}$), exhibits an intrinsic multi-orbital electronic structure involving both Ni-$3d_{x2-y2}$ and Ni-$3d_{3z2-r2}$ orbitals [25–30]. In particular, enhanced interlayer coupling mediated by the Ni-$3d_{3z2-r2}$ and apical O-$2p_z$ orbitals has been proposed as a key ingredient for the pressure-induced superconductivity. The resulting electronic and pairing states are therefore highly sensitive to the underlying lattice geometry, including the Ni–O–Ni bond angles, interlayer spacing, and local octahedral distortions [1,2]. Establishing the intrinsic crystal structure under pressure is essential not only for elucidating the microscopic origin of superconductivity in this system but also for guiding the design of nickelate materials with enhanced functionalities.

A key complication is that even the ambient-pressure structure of $La_3Ni_2O_7$ has only recently been clarified. Although early diffraction studies assigned $La_3Ni_2O_7$ to centrosymmetric orthorhombic structures, initially *Fmmm* and subsequently the lower-symmetry *Amam* (Fig. 1a) space group [1,2,16,31,32], recent high-brilliance synchrotron single-crystal X-ray diffraction measurements revealed a polar *Am*2*m* structure at ambient pressure (Fig. 1b) [33]. This polarity originates from the combined effect of charge order, which renders the two Ni sites crystallographically inequivalent, and the tilting of the $NiO_6$ octahedra. By analogy with perovskite-type rare-earth nickelates, the observed charge order may be more appropriately viewed as bond disproportionation driven by modulated Ni–O covalency, where the electronic states are better described by $3d^8$ and $3d^8\underline{L}$-like configurations ($\underline{L}$ denotes a ligand hole) rather than conventional *d*-electron charge ordering [34]. However, how the polar and charge-ordered states evolve under pressure toward the superconducting phase remains unresolved.

Furthermore, previous high-pressure studies have reported conflicting structural models for superconducting $La_3Ni_2O_7$. Early high-pressure diffraction experiments proposed an orthorhombic *Fmmm* structure characterized by linear interlayer Ni–O–Ni bonds [1], whereas subsequent X-ray and optical measurements revealed a more intricate pressure-induced structural evolution and provided evidence for a tetragonal *I*4/*mmm* structure at higher pressures (Fig. 1c) [2–5,35]. Because these proposed crystal symmetries differ substantially in lattice geometry and local coordination environment, identifying the intrinsic high-pressure structure has become a central issue in understanding the superconductivity in $La_3Ni_2O_7$.

Resolving this issue requires eliminating extrinsic effects that can obscure the intrinsic structure. Ruddlesden–Popper nickelates are highly sensitive to oxygen non-stoichiometry and the associated polymorphism [15,18,36–38], while powder diffraction measurements under non-hydrostatic pressure conditions can further complicate the structural determination through strain effects and limited symmetry resolution [1–3,11,12]. The importance of maintaining hydrostatic conditions in high-pressure crystallographic studies has been widely recognized [39], particularly for systems with competing structural instabilities. A definitive structural phase diagram therefore requires stoichiometric, polymorph-free single crystals studied under strictly hydrostatic conditions.

Here, we establish the intrinsic pressure–temperature structural phase diagram of stoichiometric $La_3Ni_2O_7$ using synchrotron single-crystal X-ray diffraction under hydrostatic helium pressure conditions. We find that the polar *Am*2*m* structure with charge order persists at low pressures and undergoes a direct transition to the tetragonal *I*4/*mmm* phase near 10 GPa, coincident with the emergence of bulk superconductivity. These results identify the intrinsic crystallographic pathway into the superconducting state. Our findings resolve a central structural ambiguity in $La_3Ni_2O_7$ and establish a structural basis for understanding nickelate high-$T_C$ superconductivity.

## Results and discussion

### High-pressure X-ray diffraction experiments

High-pressure X-ray diffraction measurements were performed on Crystal 1 (50 × 50 × 20 μm$^3$), the same stoichiometric crystal whose ambient-pressure structure was previously identified as the polar *Am2m* phase with charge order and no detectable oxygen deficiency [40]. In addition, Crystal 2 (30 × 20 × 10 μm$^3$) from the same growth batch was used for complementary measurements, as described later. The measurement points and pressure–temperature paths explored in this study are summarized in Figs. 1d and S1. The experimental geometry of the diamond anvil cell (DAC) setup is shown in Fig. 2. To sensitively probe lattice changes associated with the orthorhombic-to-tetragonal transition, X-rays were incident parallel to the *c*-axis, allowing selective access primarily to the in-plane (*ab*-plane) structural information. Diffraction data were collected over an angular range of $\omega$ = −30° to +30°.

For Crystal 1, cooling from 300 K to 9 K at 0.2 GPa resulted in no detectable changes in the diffraction pattern. The data collected at 0.2 GPa and 9 K, shown in Fig. 3a, remain consistent with an orthorhombic lattice, with all observed reflections satisfying the *A*-centering reflection condition (*hkl*: *k*+*l* = 2*n*). Focusing on the $1\,\bar{3}\,\bar{1}_o$ reflection highlighted by the pink circle in Fig. 3a, a weak additional reflection is observed at a slightly higher $2\theta$ position, which corresponds to the $3\,1\,\bar{1}_o$ reflection from a 90°-rotated orthorhombic twin domain. Although anomalies in the electrical resistivity have been reported near 150 K and 100 K at ambient pressure [40], no corresponding symmetry change or superlattice modulation was detected in the present single-crystal X-ray diffraction measurements, suggesting that these anomalies are primarily electronic or magnetic in origin. This observation is also consistent with previous low-temperature single-crystal X-ray diffraction studies performed at ambient pressure [33].

Even at 7.0 GPa and 9 K, peak splitting associated with orthorhombic domains is still observed, although its magnitude is substantially reduced (Fig. 3b). Furthermore, reflections violating the extinction condition of the previously reported *F*-centered lattice (*Fmmm*) [1–3] are clearly observed. These include reflections with mixed odd and even Miller indices, such as $\bar{2}\,5\,\overline{13}_o$. In contrast, all reflections satisfy the *A*-centering reflection condition (*hkl*: *k*+*l* = 2*n*). These results demonstrate that the crystal retains an *A*-centered orthorhombic structure.

Upon further compression at 9 K, a pronounced change in the diffraction pattern emerges. At 13.7 GPa, the weak reflections associated with orthorhombic domains disappear, and the diffraction peaks merge into single peaks (Fig. 3c). This diffraction pattern can be consistently indexed by an *I*-centered tetragonal lattice with the transformation $a_t = (a_o + b_o)/\sqrt{2}$, $b_t = (-a_o + b_o)/\sqrt{2}$, and $c_t = c_o$ (inset of Fig. 3c), where the subscripts t and o denote the tetragonal and orthorhombic phases, respectively. All observed reflections satisfy the *I*-centering reflection condition (*hkl*: *h*+*k*+*l* = 2*n*). The pressure dependence of the lattice parameters at 9 K is shown in Fig. 3e. These observations demonstrate a structural phase transition from an *A*-centered orthorhombic phase to an *I*-centered tetragonal phase. The latter remains stable up to at least 17.7 GPa (Fig. 3d). Moreover, this structure persists not only in the superconducting region but also in the higher-temperature regime where superconductivity is suppressed, indicating that the tetragonal symmetry is a robust structural feature under high pressure.

Importantly, the single crystals investigated here were taken from the same growth batch as those exhibiting bulk superconductivity with $T_C^{\mathrm{onset}}$ ≈ 68 K at pressures above 10 GPa in electrical transport measurements [40]. As summarized in Fig. 1d, the pressure–temperature region where superconductivity emerges coincides with the region where the *I*-centered tetragonal structure is stable. This behavior contrasts with previously proposed structural phase diagrams that associated superconductivity with an orthorhombic *Fmmm* phase [1–3].

### Crystal structure in the superconducting state

The structural characteristics of the candidate crystal structures proposed for bilayer $La_3Ni_2O_7$ under ambient and high pressures are summarized in Table 1. Previous X-ray diffraction experiments and theoretical calculations have suggested that the *I*4/*mmm* structure is stabilized under high pressure [2–5], consistent with the present results. In the lower-symmetry *Am*2*m* structure at ambient pressure, the $NiO_6$ octahedra exhibit finite tilting (Glazer notation: $a^-a^-c^0$ [41,42]), causing the Ni–O–Ni bond angle along the *c*-axis to deviate from 180°. In contrast, the *I*4/*mmm* structure is characterized by the complete suppression of octahedral tilting ($a^0a^0c^0$), resulting in linear interlayer Ni–O–Ni bonds and modified Ni–O orbital hybridization. This structure provides the crystallographic framework within which superconductivity emerges.

### Relationship between charge order and superconductivity

In the pressure range just below the onset of superconductivity ($P$ ≈ 10 GPa), it remains unclear from the Crystal 1 measurements whether the charge order observed at ambient pressure persists or is suppressed. This is because the high-pressure diffraction geometry used for Crystal 1 did not provide access to the *h*0*l* plane, preventing discrimination between the *Am*2*m* and *Amam* structures. Since these two structures differ only in the presence or absence of the *a*-glide symmetry associated with charge order, whereas both retain the same $NiO_6$ octahedral tilt pattern, resolving this symmetry distinction is essential for understanding the emergence of superconductivity.

To address this issue, complementary high-pressure single-crystal X-ray diffraction measurements were performed on Crystal 2 at 300 K using a geometry that provides direct access to the *h*0*l* plane (Fig. 2b). The absence of the cryostat for the room-temperature measurement greatly reduced the background signal, enabling the clear detection of the 3 0 $0_o$ reflection at 0.2 GPa (Fig. 4a). This reflection satisfies the *A*-centering reflection condition (*hkl*: $k+l = 2n$) but is forbidden in the *Amam* structure by the *a*-glide extinction condition (*h*0*l*: $h = 2n+1$; see also Fig. S2). Although its intensity is more than four orders of magnitude weaker than that of the fundamental Bragg reflections, the reflection is clearly observed, confirming the *Am*2*m* structure in the low-pressure phase.

Upon compression, the lattice parameters exhibit an orthorhombic-to-tetragonal transition near 10 GPa (Figs. 4b–4d), similar to that observed at 9 K (Fig. 3e). Remarkably, the integrated intensity of the 3 0 $0_o$ reflection decreases approximately linearly with pressure (Fig. 4e) and disappears at the same pressure where the orthorhombic distortion ($a_o/b_o < 1$) vanishes (Fig. 4d). This coincidence demonstrates a direct structural transition from the charge-ordered *Am*2*m* phase to the tetragonal *I*4/*mmm* phase, without an intermediate *Amam* phase. Although both the lattice parameters and the 3 0 $0_o$ intensity evolve continuously with pressure, the direct *Am*2*m*-to-*I*4/*mmm* transformation is more consistent with a first-order structural transition than with a continuous symmetry-breaking transition. These results provide compelling structural evidence that the suppression of both $NiO_6$ tilting and charge order accompanies the emergence of superconductivity.

### Absence of the *Fmmm* phase under high pressure

In the present high-pressure X-ray diffraction measurements using a high-quality stoichiometric single crystal and a helium pressure medium, we found no evidence for the orthorhombic *Fmmm* phase (Fig. 3b) reported in previous structural studies [1–3]. This discrepancy likely arises from variations in sample characteristics and experimental conditions. Ruddlesden–Popper nickelates are prone to polymorphism, often exhibiting closely related structures with similar lattice parameters [2,36–38]. $La_3Ni_2O_7$, for example, crystallizes in either orthorhombic or monoclinic polymorphs depending on the synthesis conditions [1,13,18,36,37], making structural determination from powder diffraction alone particularly challenging under pressure. In addition, differences in hydrostaticity may affect the relative stability of competing structural distortions [39]. These considerations highlight that single-crystal measurements under highly hydrostatic conditions are essential for establishing the intrinsic crystal structure of bilayer $La_3Ni_2O_7$ at high pressure.

## Conclusion

Our hydrostatic single-crystal X-ray diffraction measurements resolve the central crystallographic ambiguity surrounding superconducting bilayer $La_3Ni_2O_7$ and establish its intrinsic pressure–temperature structural phase diagram. We find that the polar *Am*2*m* phase persists at low pressure and transforms directly into a tetragonal *I*4/*mmm* phase near 10 GPa, coincident with the onset of bulk superconductivity. The previously proposed orthorhombic *Amam* and *Fmmm* phases are not observed throughout the investigated pressure–temperature range, up to 19.5 GPa and 9–300 K. The pressure-induced transition to the tilt-free *I*4/*mmm* structure establishes a high-symmetry crystallographic framework with linear interlayer Ni–O–Ni bonds for the superconducting state. At the same time, our results indicate that this structural transition alone is unlikely to be sufficient for superconductivity, as the tetragonal phase persists over a broader pressure–temperature range than the superconducting state itself. The present work shifts the focus from identifying the crystal structure of superconducting $La_3Ni_2O_7$ to understanding which additional electronic or magnetic degrees of freedom are responsible for the emergence of superconductivity within this structural framework [43,44]. Comprehensive hydrostatic single-crystal diffraction studies across the nickelate family will be crucial for disentangling the structural prerequisites for superconductivity from the microscopic interactions that ultimately drive it.

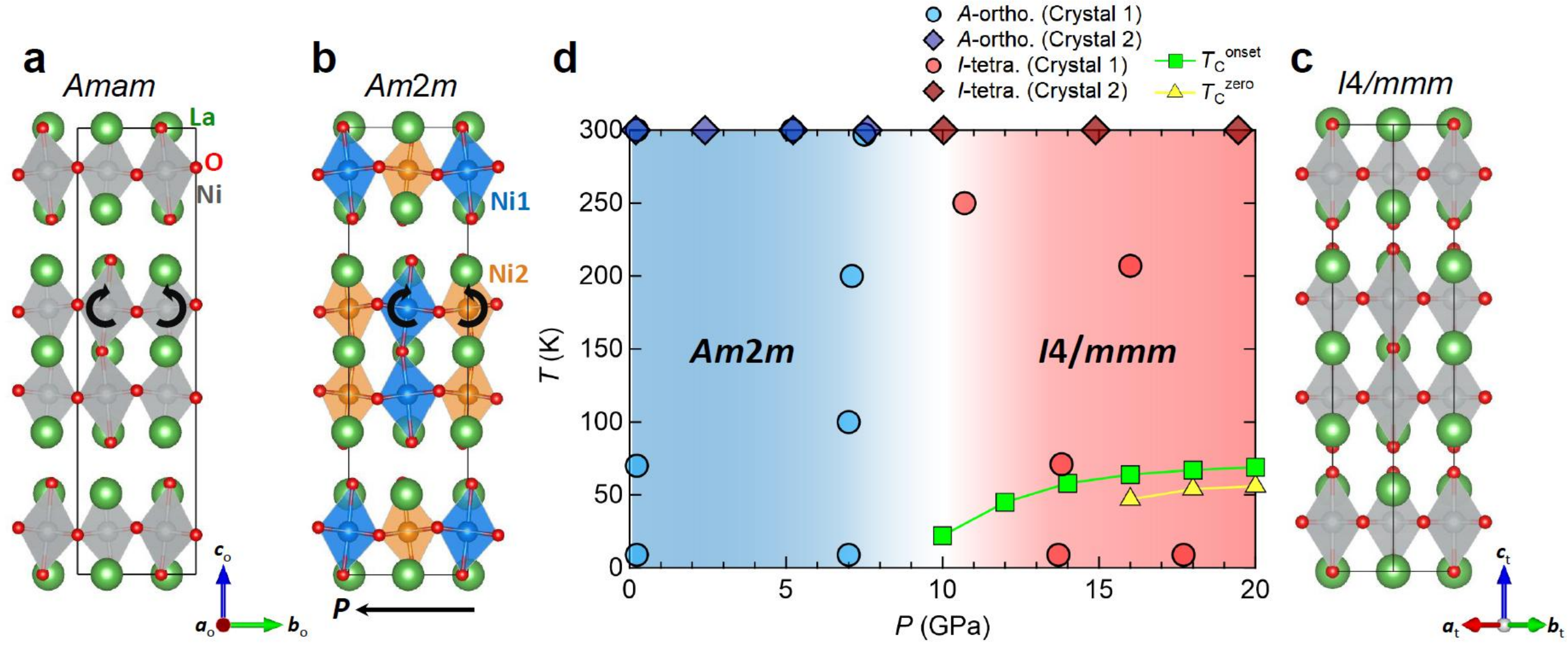

**Fig. 1. Crystal structures and pressure–temperature phase diagram of bilayer $La_3Ni_2O_7$. a**, Crystal structure with the orthorhombic *Amam* space group, in which all Ni atoms occupy a single crystallographic site and the $NiO_6$ octahedra exhibit finite tilting. The present single-crystal X-ray diffraction measurements rule out this structure for bilayer $La_3Ni_2O_7$ at both ambient and high pressures. **b**, Orthorhombic *Am*2*m* structure at ambient and low pressures. Blue and orange octahedra represent the large- and small-volume $NiO_6$ octahedra, respectively. The combined effects of charge order between inequivalent Ni1 and Ni2 sites and octahedral tilting give rise to polarity along the $b_o$ axis. **c**, Tetragonal *I*4/*mmm* structure under high pressure, in which both charge order and octahedral tilting are absent. **d**, Pressure–temperature phase diagram of stoichiometric $La_3Ni_2O_7$ single crystals. Circles denote the measurement points for Crystal 1 at low temperatures, whereas diamonds denote those for Crystal 2 at 300 K. Blue and red symbols correspond to the *Am*2*m* and *I*4/*mmm* phases, respectively. The onset ($T_C^{\mathrm{onset}}$) and zero-resistance ($T_C^{\mathrm{zero}}$) superconducting transition temperatures, taken from Ref. [40], are indicated by light-green squares and yellow triangles, respectively.

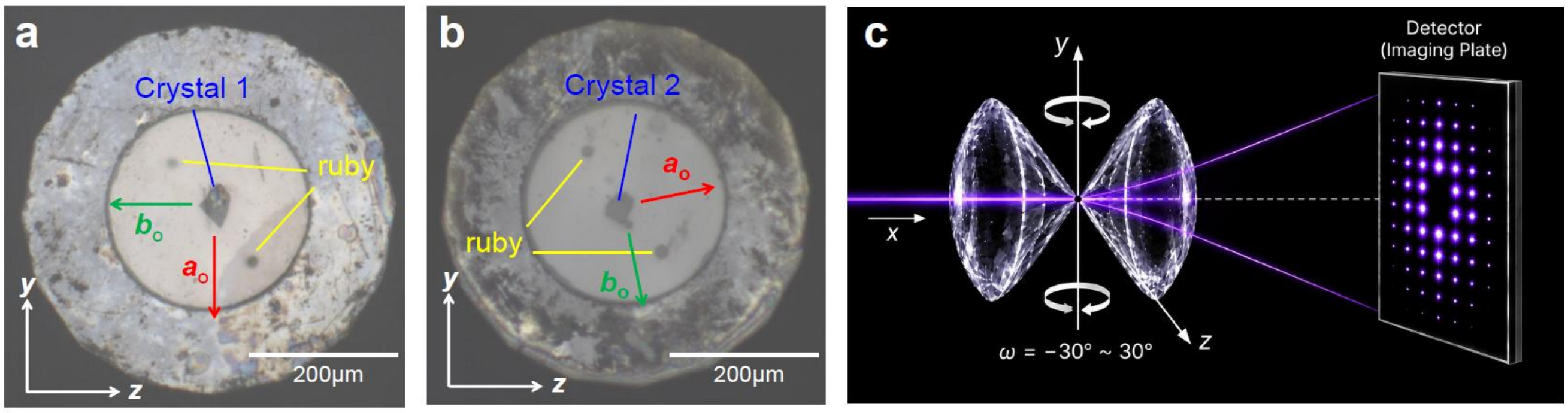

**Fig. 2. DAC and X-ray diffraction conditions under pressure. a**,**b**, Photographs of Crystal 1 and Crystal 2, respectively, loaded into the gasket hole of the DAC with different crystallographic orientations after helium gas loading. The pressures were 0.20 and 0.21 GPa, respectively. **c**, Schematic illustration of the high-pressure single-crystal X-ray diffraction setup using a DAC.

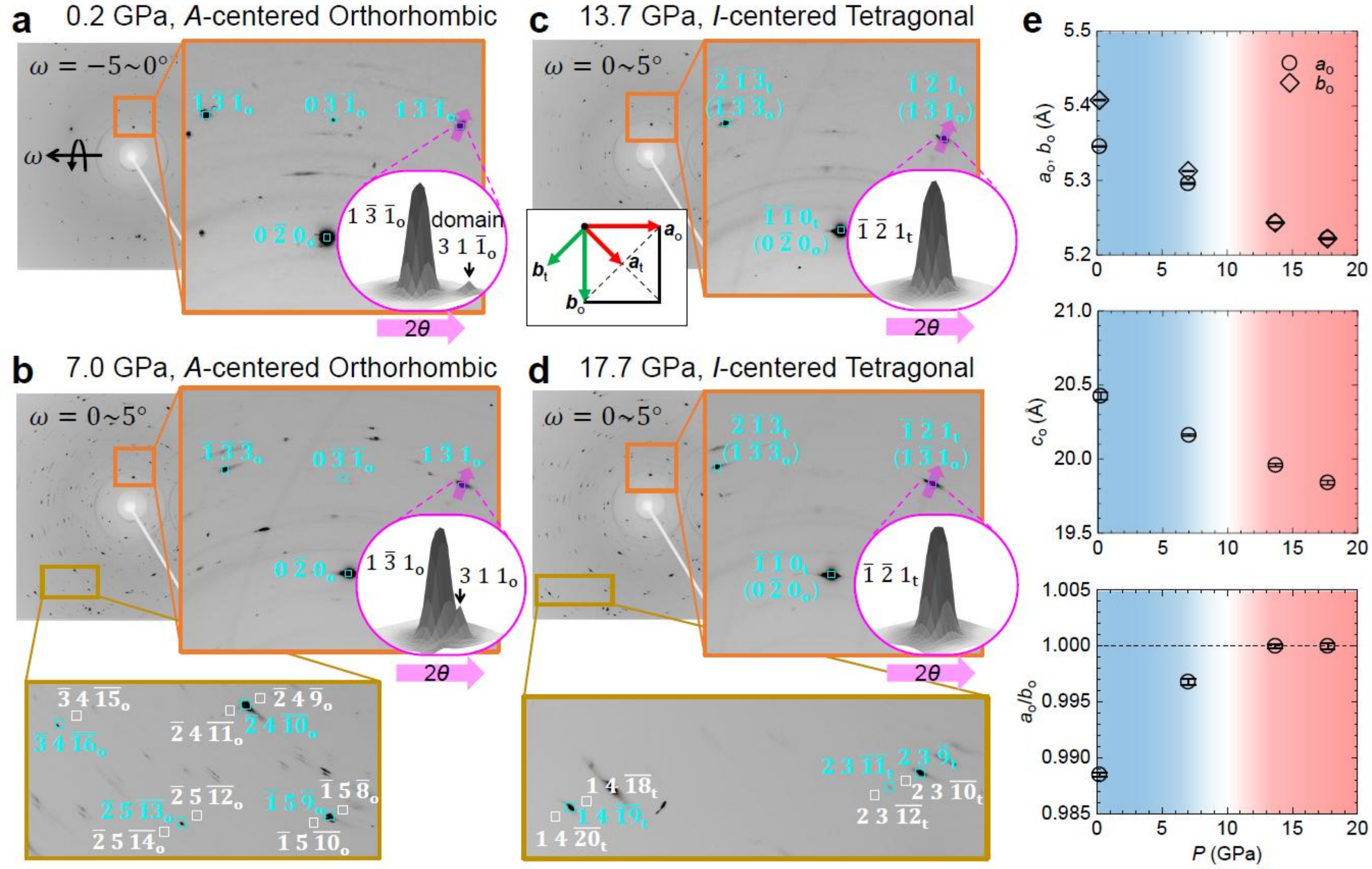

**Fig. 3. High-pressure X-ray diffraction results for Crystal 1 at 9 K. a-d**, X-ray diffraction pattern collected at 0.2, 7.0, 13.7, and 17.7 GPa, respectively. Reflections satisfying the centering conditions of the respective lattices are indicated by blue squares. In the enlarged views shown in the lower panels of (**b**) and (**d**), reflection positions corresponding to the extinction conditions are indicated by white squares. The three-dimensional intensity profile of the Bragg peaks is shown within the pink circle in the inset. Note that the crystal orientation changes slightly during the application and release of pressure. The datasets were collected in the order of (**a**), (**c**), (**d**), and (**b**). After passing through the pressure-induced tetragonal phase, the crystal recovered the orthorhombic phase with a larger number of domains, resulting in the appearance of more orthorhombic-domain-related reflections in (**b**) than in (**a**). **e**, Pressure dependence of the lattice parameters $a_o$, $b_o$, $c_o$, and the $a_o/b_o$ ratio. For consistency, all parameters are expressed using the orthorhombic setting throughout the entire pressure range.

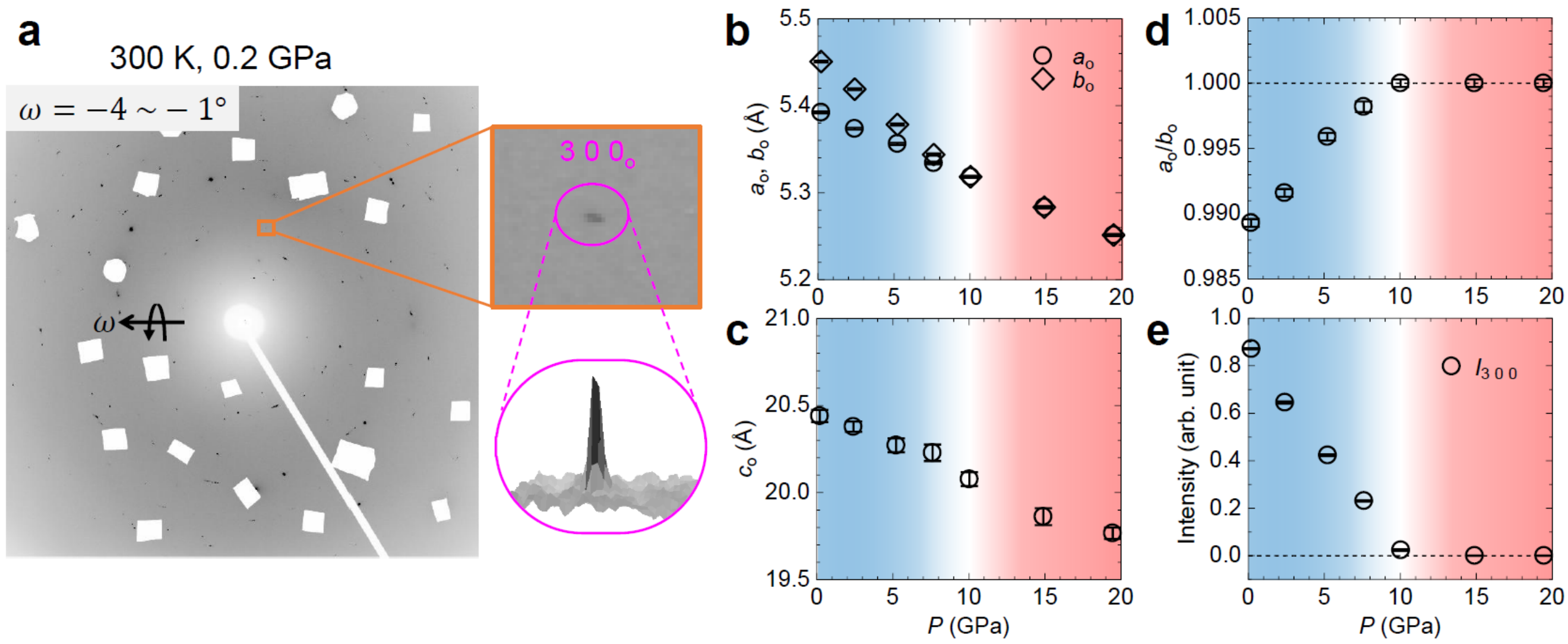

**Fig. 4. High-pressure X-ray diffraction results for Crystal 2 at 300 K. a**, X-ray diffraction pattern collected at 0.2 GPa. White squares indicate Pb plates placed over the diamond Bragg peak positions on the detector to protect it from intense diamond reflections, enabling long-exposure measurements of weak diffraction signals. The enlarged 3 0 $0_o$ reflection, highlighted by pink circles, violates the extinction condition associated with the *a*-glide plane perpendicular to the $b_o$ axis ($h0l$: $h$ = $2n$+1). **b-d**, Pressure dependence of the lattice parameters $a_o$, $b_o$, $c_o$, and the $a_o/b_o$ ratio. **e**, Pressure dependence of the integrated intensity of the 3 0 $0_o$ reflection.

Table 1. Structural characteristics of the candidate crystal structures for bilayer $La_3Ni_2O_7$. Octahedral tilt patterns are described using the Glazer notation [41]. Here, $a^-a^-c^0$ denotes out-of-phase octahedral tilting about the pseudocubic *x*- and *y*-axes, whereas $a^0a^0c^0$ indicates the absence of octahedral tilting.

| Space group | Lattice | In-plane axes | $NiO_6$ tilt | Ni charge order | Polarity |
|---|---|---|---|---|---|
| *I*4/*mmm* | $\mathbf{a}_t$, $\mathbf{b}_t$, $\mathbf{c}_t$ | $a_t = b_t$ | $a^0a^0c^0$ | × | × |
| *Fmmm* | $\mathbf{a}_o = \mathbf{a}_t - \mathbf{b}_t$,<br>$\mathbf{b}_o = \mathbf{a}_t + \mathbf{b}_t$,<br>$\mathbf{c}_o = \mathbf{c}_t$ | $a_o \neq b_o$ | $a^0a^0c^0$ | × | × |
| *Amam* | Same as above | $a_o \neq b_o$ | $a^-a^-c^0$ | × | × |
| *Am*2*m* | Same as above | $a_o \neq b_o$ | $a^-a^-c^0$ | ○ | ○ |

## Methods

Sample preparation

Single crystals of bilayer $La_3Ni_2O_7$ were grown under high-pressure conditions, as described in Ref. [40].

X-ray diffraction under high pressure

High-pressure X-ray diffraction measurements were performed at the BL10XU beamline [45] of SPring-8, where a two-dimensional imaging plate detector is installed. A DAC with a culet size of 500 μm was used for high-pressure experiments. The same single crystal used for the ambient-pressure measurements was loaded into a 260-μm-diameter hole in a rhenium gasket, together with helium precompressed to 200 MPa using a high-pressure gas-loading system at SPring-8. Pressure was determined from the ruby fluorescence using the scale of Shen *et al*. [46], with a temperature correction based on Ragan *et al.* [47]. X-rays with an energy of 30 keV and a beam diameter of 60 μm were used. Temperature was controlled using a cryostat, and pressure was tuned using a helium gas membrane system.

## Acknowledgements

We thank N. Ogawa for providing the optical setup used for preparing gaskets, and R. Arita for fruitful discussions. This work was supported by JSPS KAKENHI (Grant Nos. 24K17006, 24H01644, 26H01288, 26H00626, and 26K00649), JST FOREST (Grant No. JPMJFR2362), JST CREST (Grant No. JPMJCR26X8), JST PRESTO (Grant No. JPMJPR2597). H.Y. was supported by Iwadare Scholarship Foundation. The synchrotron radiation experiments were performed at SPring-8 with the approval of the Japan Synchrotron Radiation Research Institute (JASRI) (Proposal No. 2026A1163, and 2026A1308).

## Author contributions

S.K., Y.F. designed and coordinated the study. Z.L., M.K., Y.Ta. grew the single crystals. M.S., C.T., Y.F. prepared the DAC. M.S., T.H., S.K., Y.F., H.Y., S.Y., Z.L., N.H., H.K., performed the high-pressure X-ray diffraction experiment. T.H., S.K. analyzed the high-pressure X-ray diffraction data. S.K., Y.F. wrote the manuscript under the supervision of Y.W., Y.To., Y.Ta., T.A. All authors discussed the experimental results and contributed to the manuscript.

## Competing interests

The authors declare no competing interests.

# Single-crystal structural phase diagram of stoichiometric bilayer nickelate $La_3Ni_2O_7$ under hydrostatic pressure

Misaki Sasaki[1], Zhehong Liu[2], Takeshi Hara[3], Shunsuke Kitou[4], Markus Kriener[2], Haruto Yoshimochi[2], Shion Yamada[4], Chieko Terakura[2], Naohisa Hirao[5], Hirokazu Kadobayashi[5], Yusuke Wakabayashi[3], Yoshinori Tokura[1,2,6], Yasujiro Taguchi[2], Taka-hisa Arima[2,4], Yukako Fujishiro[1,2,7,8]

[1]*Department of Applied Physics, The University of Tokyo, Tokyo 113-8654 Japan*
[2]*RIKEN Center for Emergent Matter Science (CEMS), Wako 351-0198, Japan*
[3]*Department of Physics, Graduate School of Science, Tohoku University, Sendai 980-8578, Japan*
[4]*Department of Advanced Materials Science, The University of Tokyo, Kashiwa 277-8561, Japan*
[5]*Japan Synchrotron Radiation Research Institute (JASRI), SPring-8, Hyogo 679-5198, Japan*
[6]*Tokyo College, The University of Tokyo, Tokyo 113-8654 Japan*
[7]*Institute of Engineering Innovation, The University of Tokyo, Tokyo 113-8656, Japan*
[8] *RIKEN Pioneering Research Institute (PRI), Saitama 351-0198, Japan*

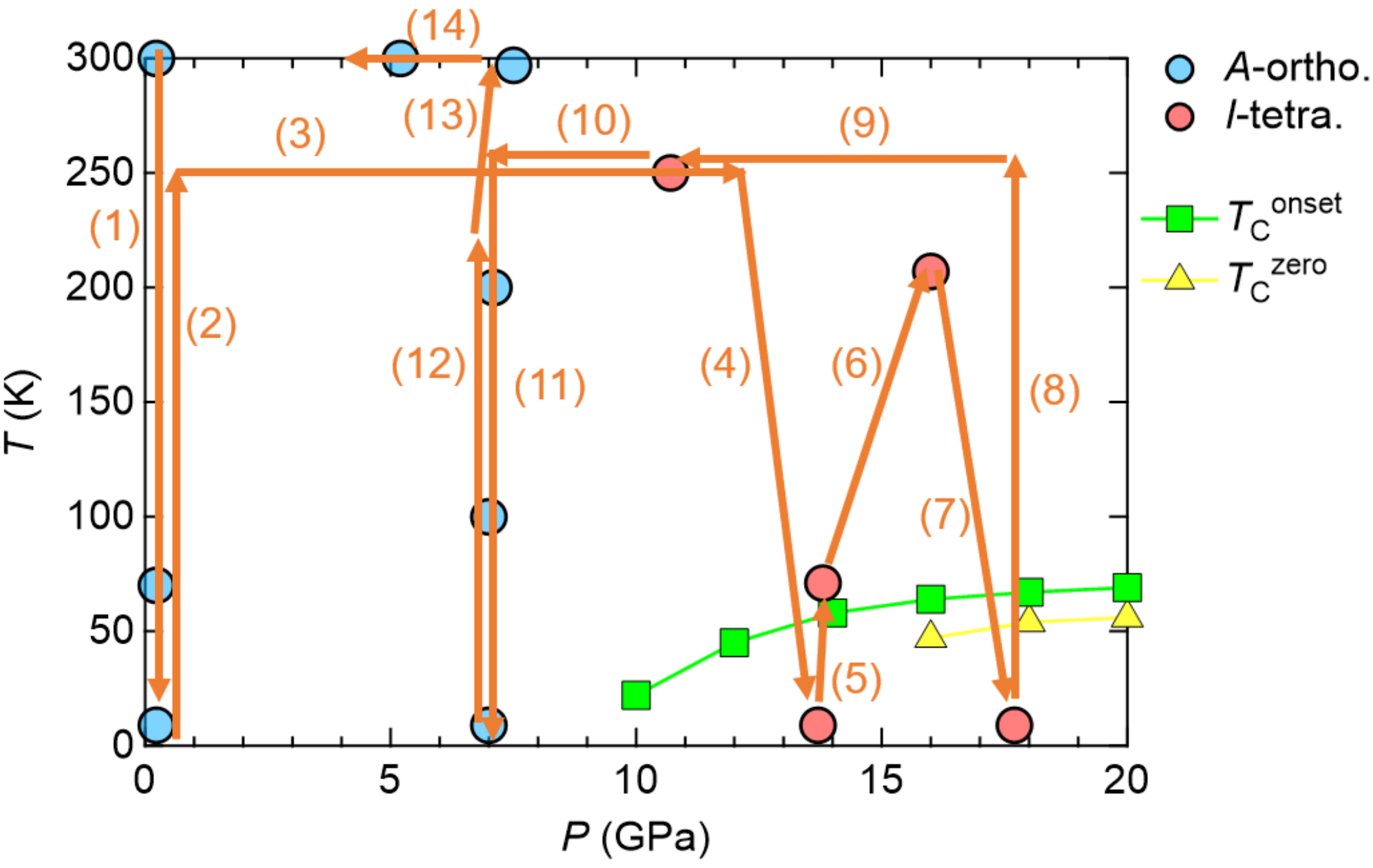

Figure S1. X-ray diffraction measurement path on the pressure–temperature phase diagram of Crystal 1. Circles denote the measurement points. The numbers in parentheses attached to the arrows indicate the chronological sequence of temperature and pressure changes during the measurements. The onset ($T_{\mathrm{C}}^{\mathrm{onset}}$) and zero-resistance ($T_{\mathrm{C}}^{\mathrm{zero}}$) transition temperatures of superconductivity, taken from Ref. [S1], are indicated by light-green squares and yellow triangles, respectively.

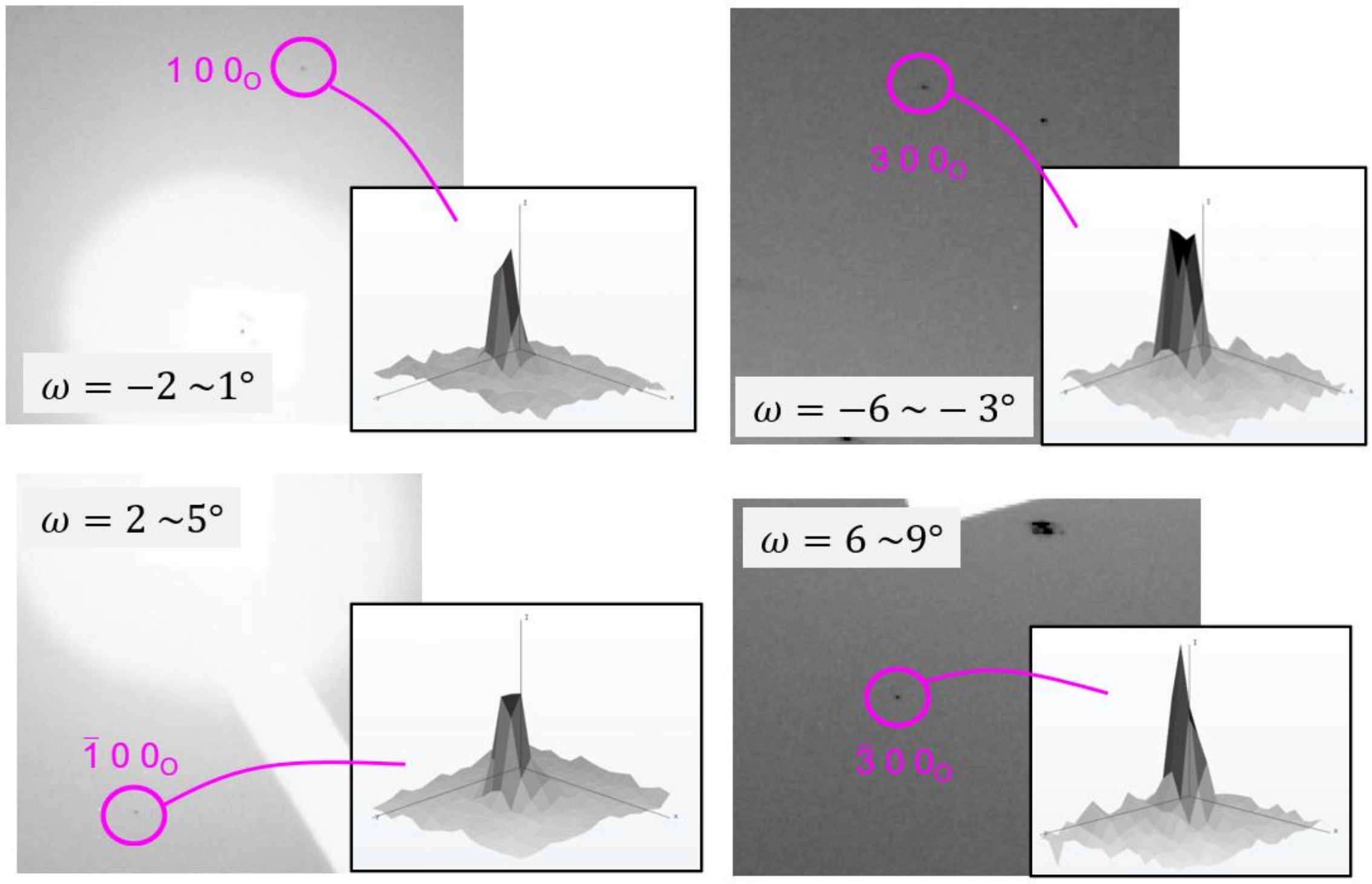

Figure S2. X-ray diffraction patterns for Crystal 2 at 0.2 GPa and 300 K in the orthorhombic *Am*2*m* phase. Several reflections satisfying *h*0*l*: *h* = 2*n*+1 are observed. All of these reflections disappear above 10 GPa. Their systematic appearance at low pressure and simultaneous disappearance above the structural transition demonstrate that these reflections are intrinsic violations of the *a*-glide extinction condition of the *Amam* structure, rather than artifacts arising from multiple scattering.